\documentclass[10pt,conference]{IEEEtran}
\IEEEoverridecommandlockouts
\usepackage{cite}
\usepackage{amsmath,amssymb,amsfonts}
\usepackage[ruled, linesnumbered]{algorithm2e}
\usepackage{graphicx}
\usepackage{textcomp}
\usepackage[table]{xcolor}
\usepackage{booktabs}
\usepackage{listings}
\usepackage{url}
\usepackage{balance}
\usepackage{multirow}

\begin{document}

\title{Automated Test-Case Generation for REST APIs Using Model Inference Search Heuristic}


\author{\IEEEauthorblockN{Clinton Cao}
\IEEEauthorblockA{\textit{Delft University of Technology}\\
Delft, the Netherlands \\
c.s.cao@tudelft.nl}
\and
\IEEEauthorblockN{Annibale Panichella}
\IEEEauthorblockA{\textit{Delft University of Technology}\\
Delft, the Netherlands \\
a.panichella@tudelft.nl}
\and
\IEEEauthorblockN{ Sicco Verwer}
\IEEEauthorblockA{\textit{Delft University of Technology}\\
Delft, the Netherlands \\
s.e.verwer@tudelft.nl}
}

\maketitle

\begin{abstract}
The rising popularity of the microservice architectural style has led to a growing demand for automated testing approaches tailored to these systems. EvoMaster is a state-of-the-art tool that uses Evolutionary Algorithms (EAs) to automatically generate test cases for microservices' REST APIs. One limitation of these EAs is the use of unit-level search heuristics, such as branch distances, which focus on fine-grained code coverage and may not effectively capture the complex, interconnected behaviors characteristic of system-level testing.
To address this limitation, we propose a new search heuristic (MISH) that uses real-time automaton learning to guide the test case generation process. 
We capture the sequential call patterns exhibited by a test case by learning an automaton from the stream of log events outputted by different microservices within the same system. Therefore, MISH learns a representation of the system-wide behavior, allowing us to define the fitness of a test case based on the path it traverses within the inferred automaton. 
We empirically evaluate MISH's effectiveness on six real-world benchmark microservice applications and compare it against a state-of-the-art technique, MOSA, for testing REST APIs. Our evaluation shows promising results for using MISH to guide the automated test case generation within EvoMaster.
\end{abstract}


\section{Introduction}
In recent years, the microservice architectural style has gained significant traction among companies developing modern (web) applications. This shift is largely driven by the benefits provided by this architectural style: it enables developers to develop, maintain, and deploy the components (known as microservices) of their applications independently. Microservices interact with each other via REST APIs, and a request by a client may involve a sequence of REST API calls between microservices to get the corresponding response. Despite the benefits, microservices are notoriously known to be hard to debug once a fault has occurred due to their distributed nature. Different studies have shown that developers usually spend days debugging a fault, making it a time-consuming process~\cite{Zhou2018_fault_analysis,Lenarduzzi_Panichella_2021_serverless}. 

As with any software system, writing system-level tests could mitigate potential faults after deployment. 
However, manually crafting these tests for microservices is not only labor-intensive but also prone to incomplete coverage as developers might not cover all possible cases that could occur in the system due to their complexity~\cite{Lenarduzzi_Panichella_2021_serverless,Waseem_2020_Testing_Microservices,Waseem_2021_Design,Rafi_2012_Benefits_Automated_Testing}. These challenges have lead to the growing need of tools to generate test cases for microservices automatically. EvoMaster is a state-of-the-art tool that uses Evolutionary algorithms (EAs)~\cite{Arcuri_2019_Restful} to create test cases for microservice applications with web APIs. 
Specifically, the EA continuously generates and evolves a pool of test cases that can be executed against the system under test (SUT). Each test case represents a sequence of REST API calls for a particular type of resource. Test cases are evolved based on a set of objectives (e.g., line coverage, branch distances, etc.) used to define the fitness of the test cases. Typically, randomized operations such as mutation and crossover are used to evolve the test cases. Furthermore, different (randomized) operations are used to select test cases to keep within the pool.  

Despite their success, existing EA-based approaches rely heavily on unit-level search heuristics, such as branch distances, to guide the generation of test cases. Branch distance, for example, measures how close a test case is to cover a specific branch in the code using Korel's rules~\cite{korel1990automated} for branching conditions. While effective for maximizing fine-grained code coverage within individual units or services, these heuristics provide no guidance on system-level behaviors. Specifically, they lack awareness of how microservices interact and how execution flows across services. This narrow focus on unit-level objectives can result in test cases that fail to explore distributed systems' intricate, interconnected behaviors, thereby missing faults that arise from complex inter-service interactions~\cite{Stallenberg_2021_Improving}. 


To address this limitation, we introduce the Model Inference Search Heuristic (MISH), a novel grey-box approach to guide the automatic test case generation process. Unlike white-box techniques that rely on code-level metrics such as branch distance, MISH leverages log statements printed by the SUT during test execution, which developers insert into the code. MISH continuously collects log statements output by the SUT during test execution and uses them to learn a finite state automaton (or simply a state machine) in real-time. Each test case is converted into a sequence of events derived from these logs, and the inferred state machine serves as a representation of the system's behavior. We learn a state machine using the online learning heuristic proposed by Baumgartner and Verwer for FlexFringe~\cite{Baumgartner_2023_Learning}, a framework for learning state machines. Our approach uses a single objective to define the fitness of a test case, the path traversed within the state machine. To the best of our knowledge, we are the first to apply real-time state machine learning to the task of automated test-case generation for microservices' REST APIs.

As state machines capture sequential behaviors exhibited by a SUT, we can leverage this information to capture the required sequential pattern of REST API calls and create the corresponding sequential behavior within the SUT. This way, we can preserve sequential patterns that can be used to generate test cases for complex interconnected behaviors between microservices. We believe these sequential patterns will lead to better code coverage and fault detection. 


To assess MISH's feasibility and effectiveness, we implemented a prototype of MISH within EvoMaster and tested it against six different real-world microservice applications provided by the EvoMaster Benchmark (EMB) dataset~\cite{Arcuri_2023_EMB}. Additionally, we compared MISH against the state-of-the-art many-objective algorithm, MOSA, for testing REST APIs. An implementation of MOSA has already been provided within EvoMaster. Our evaluation results indicate that MISH performs comparably to MOSA and can even surpass MOSA in specific scenarios while utilizing only a single objective to evaluate fitness for test cases. Moreover, our results indicate that MISH is more efficient than MOSA in covering test targets at the start of the search. This suggests that MISH captures essential patterns that could lead to quicker discovery of unseen targets within the SUT.

Our main contributions are summarized as follows:
\begin{itemize}
    \item A novel approach, MISH, that uses real-time automaton learning as a search heuristic to generate system-level test cases for REST APIs automatically.
    \item A prototype of MISH implemented within EvoMaster, which can be used for replication studies.
    \item A comparison of MISH against the state-of-the-art algorithm (MOSA) for system-level test-case generation on six different microservice applications.
\end{itemize}
\section{Background \& Related Work}\label{sec:background_related}
This Section describes the concepts used in this work, this would help the reader understand the methods utilized in MISH. Additionally, we list related works for each concept.

\subsection{Automated Test Case Generation using EAs}
EAs are widely applied within the field of search-based software testing to automate different software testing tasks such as test case generation~\cite{Fraser_2011_EvoSuite, panichella2015reformulating}, minimizing test suites~\cite{yoo2012regression}, and prioritizing test cases~\cite{birchler2023single}. Prior studies have shown that automating these tasks has significantly reduced the time that developers spend on testing and debugging their system~\cite{Amannejad_2014_A_Search-based,Solatani_2020_Search-Based}. Several automated software testing tools also incorporate EAs to generate test cases for various level of testing, such as EvoSuite~\cite{Fraser_2011_EvoSuite}, EvoMaster~\cite{Arcuri_2018_EvoMaster}, SUSHI~\cite{Braione_2017_Combining}, EXSYST~\cite{Gross_2012_Search-based}, and Sapienz~\cite{Alshahwan_2018_Deploying}. 

Inspired by the biological evolution process in nature, an EA starts with an initial population of randomly sampled test cases. In the context of software testing, a test case represents an individual in the population, and the statements in a test represent its genes. EA runs the test cases to determine their fitness values based on a predetermined fitness function (e.g. line coverage). The EA then generates offspring from the population of test cases using genetic operations such as crossover or mutation. The offspring are also run to determine their fitness values. Finally, EA selects the best candidates from the parent and offspring population to form the new population of test cases that survive to the next iteration of the EA. This process (evaluation, generation of offspring, and selection) is repeated until the stopping criterion is met (e.g., search budget time is exhausted). 

\subsection{Automated Testing of REST APIs}
Representation State Transfer (REST) APIs are predominantly used for communication or requesting resources from a (micro) service. These are typically done by executing (HyperText Transfer Protocol) HTTP calls towards pre-defined endpoints. With the increasing adoption of the microservice architectural style, testing REST APIs automatically has also become an important topic of research. Testing REST APIs comes in two flavors: black-box or white-box. The former does not have access to the source/byte code of the system, meaning no information on the SUT's internal structure can be used to guide the test case generation. The latter approach has access to the internal structure and can use this information to generate test cases to trigger specific parts of the SUT. 

RESTTESTGEN~\cite{Viglianisi_2020_RESTTESTGEN}, RESTest~\cite{Martin-Lopez_2021_RESTest}, and RapiTest~\cite{Felicio_2023_RapiTest} are methods developed to automatically test REST APIs using black-box heuristics. EvoMaster also provides the option to generate test cases following in black-box mode~\cite{Arcuri_2021_Automated}. These typically use the API specifications of the SUT to generate test cases. For the white-box approach, many state-of-the-art techniques are built as search algorithms within EvoMaster. The Many Independent Objective (MIO) algorithm is a state-of-the-art technique built within EvoMaster~\cite{Arcuri_2017_Many}. This algorithm treats each test target as an independent objective and tries to find a test case that covers the objective. The aim is to find a test suite that covers as many (independent) objectives as possible. MIO keeps track of a population of test cases for each objective and creates the final test suite by taking the union of the best test cases from all populations. The Many Objective Sorting Algorithm (MOSA) is another state-of-the-art technique built within EvoMaster. Proposed by Panichella et al.~\cite{panichella2015reformulating}, this algorithm treats test case generation as a multi-objective optimization problem. Instead of optimizing each objective separately, as with the MIO algorithm, all objectives are optimized simultaneously in MOSA. Unlike MIO, MOSA keeps track of a single population of test cases. The final result is a minimal test suite containing the test cases covering as many objectives as possible. 

\subsection{State Machine \& Testing REST APIs}
A state machine, also known as a finite state automaton, is a mathematical model for capturing the sequential behavior exhibited by a SUT. Formally, it is defined as a quintuple $(\Sigma, Q, q_0, \delta, F)$~\cite{Sipser_2013_introduction}, where:

\begin{itemize}
    \item $\Sigma$ is a finite alphabet
    \item $Q$ is a finite set of states
    \item $q_0 \in Q$ is a unique start state
    \item $\delta: Q \times \Sigma \to Q$ is the transition function, denoting which state the state machine will transition to based on the current state that it is in and the next input symbol $a \in \Sigma$.
    \item $F \subseteq Q$ is the set of final states
\end{itemize}

\begin{figure}
    \centering
    \includegraphics[width=0.80\columnwidth]{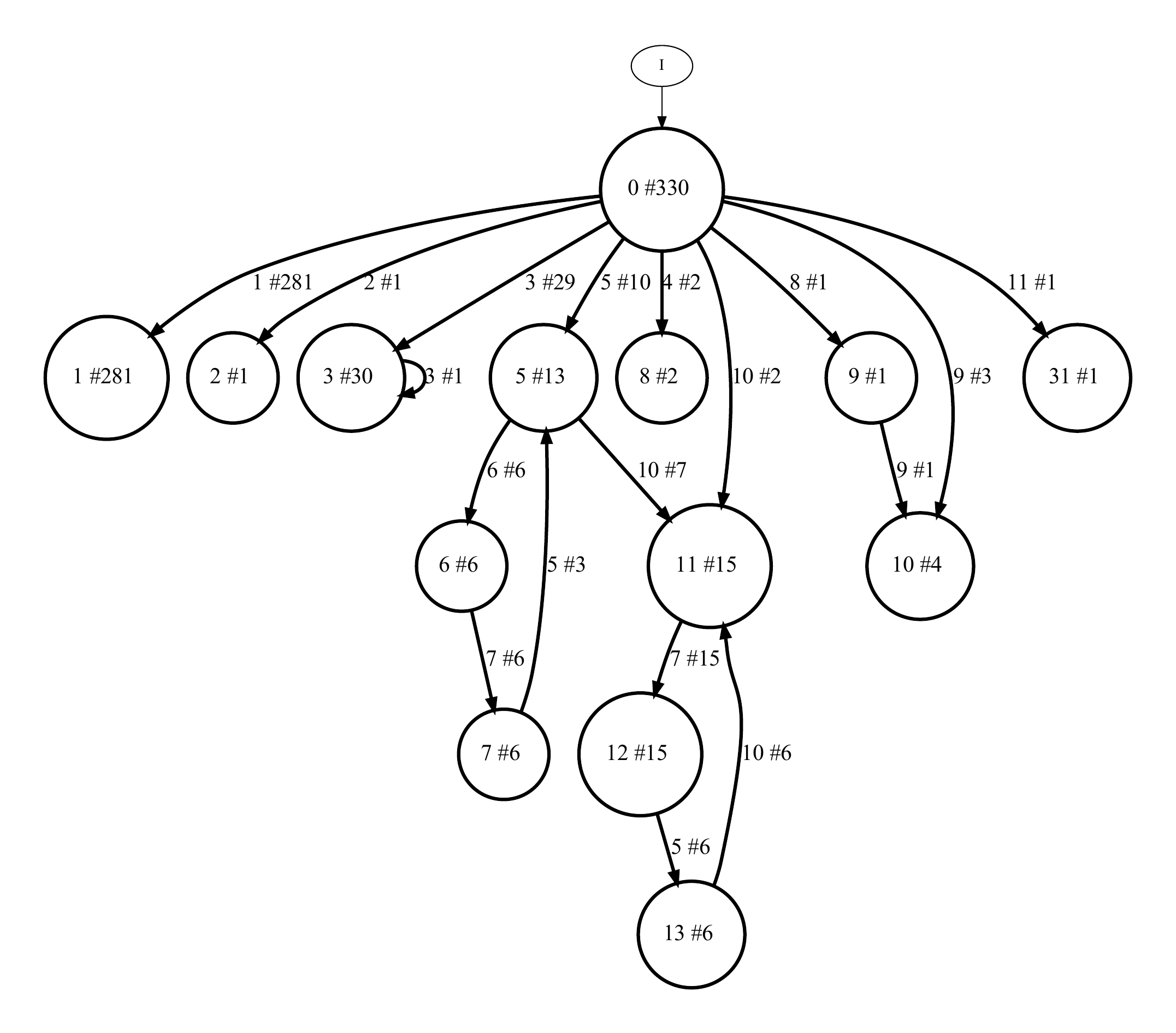}
    \caption{Example state machine learned for a SUT $S$ used in the experiments. Running the trace $t = [10,7,5,10]$, $S$ will traverse states 0, 11, 12, 13, and end in state 11. This denotes the sequential behavior $S$ exhibited for trace $t$.}
    \label{fig:example_sm_proxyprint}
\end{figure}

Given an arbitrary trace $t$ containing an arbitrary number of input symbols and a SUT $S$, the state machine can be used to understand the sequential behavior that $S$ exhibited by looking at the sequence of traversed states. In this work, we use a slightly different definition of a state machine: we do not explicitly have final states, any state within the state machine is considered a final state. Figure~\ref{fig:example_sm_proxyprint} provides an example of a state machine learned from a microservice application used in our evaluation experiments. The number after the $\#$-symbol denotes the visit frequency of a state/transition, and the number before the symbol denotes the state ID number or the transition symbol (i.e., the symbol for triggering the transition). 

There are two primary approaches to learning state machines from data: \textit{active} and \textit{passive} learning. The former uses the well-known $L^{*}$ algorithm proposed by Angluin~\cite{Angluin_1987_learning}. The algorithm continuously queries the SUT for outputs by providing different inputs. The learning process starts by collecting input-output pairs to construct a state machine. The algorithm then tries to find counterexamples to verify the correctness of the state machine. If a counterexample is found, the state machine is updated. These two steps are repeated until no more counterexamples can be found.

In contrast to the active learning approach, the passive learning approach does not need to query the SUT actively to collect data for learning. It solely collects output data of the SUT (i.e., only listening to the output) and constructs a model from the collected data. The passive learning approach uses state-merging algorithms, which were shown to be most effective for learning a model within this approach~\cite{lang_1998_results}.

Several works have shown the effectiveness of state machines in testing and verification of the behavior of different systems~\cite{Zhao_2024_AGLFuzz,Ruiter_2015_Protocol,Fiterau_2020_Analysis,Lin_2018_TABOR,Matoursek_2021_Efficient,Grov_2019_Towards}. However, the focus of these works was not on automated test-case generation for REST-APIs. Cao et al. have applied state machines on REST-APIs~\cite{Cao_2022_Learning,Cao_2024_CATMA, Cao_2024_SEQUENT} for different purposes. However, the focus of these two works was also not on automated test-case generation. Several works have specifically used state machines to generate test cases automatically for REST APIs. These approaches typically extract a state machine from the UML diagrams generated for the microservices~\cite{Pinheiro_2013_Model-Based,Yue_2011_Automated,Andrews_2005_Testing}. These approaches assume that such diagrams are available for the SUT and are up to date with the latest version of the SUT.

\section{Approach}
\begin{figure}
    \centering
    \includegraphics[width=\columnwidth]{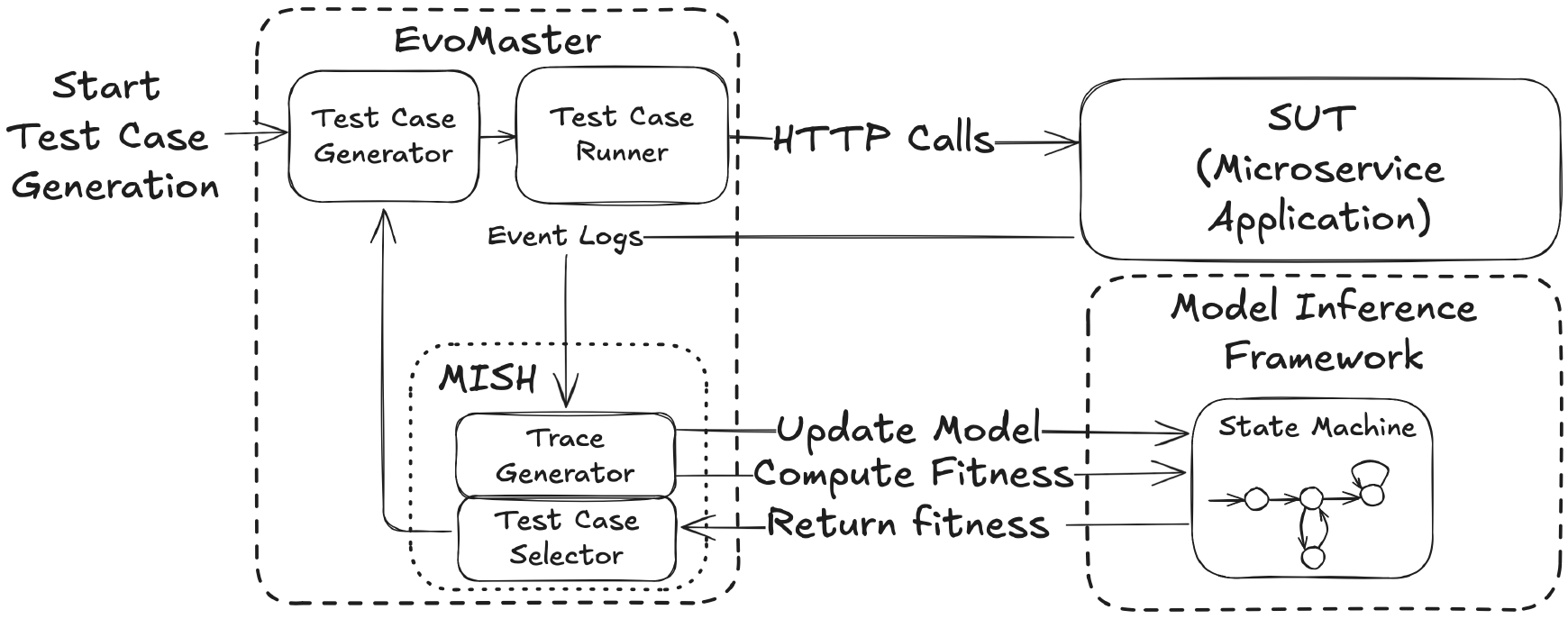}
    \caption{High-level overview of MISH's workflow.}
    \label{fig:mish_framework_diagram}
\end{figure}

Our approach continuously learns a state machine model from event logs outputted by the SUT and uses the model to define fitness for the test cases that have been executed. This section details the internal workings of our novel algorithm, MISH, 
including (1) the transformation of log events outputted by the SUT into traces, (2) the interaction between EvoMaster and the model inference framework to update a state machine with these traces continuously, (3) the computation of the fitness for each test case, and (4) the evolution of test cases for future iterations/generations. Figure~\ref{fig:mish_framework_diagram} depicts a high-level overview of MISH's workflow, and Algorithm~\ref{algo:mish} outlines the pseudo-code.

\begin{algorithm}
\LinesNumbered
\caption{MISH}\label{algo:mish}
\KwIn{Population size $S$, Fitness function $F$}
\KwOut{ A test suite $TS$}
$P \leftarrow randomPopulation(S)$\\
$runPopulation(P)$\\
$C \leftarrow collectTimeStamps(P)$\\
$T \leftarrow generateTraces(C, P)$\\
$TS \leftarrow updateArchive(P)$\\
$updateModel(T)$\\
$computeFitness(T, P, F)$\\
\While{search budget not exhausted} {
    $O \leftarrow generateOffspring(P, S)$\\
    $runPopulation(O)$\\
    $C \leftarrow collectTimeStamps(O)$\\
    $T_{O} \leftarrow generateTraces(C_{O}, O)$\\
    $updateModel(T_{O})$\\
    $computeFitness(T_{O}, O, F)$\\
    $computeFitness(T, P, F)$\\
    $P, T \leftarrow generateNewPopulation(P, O, T, T_{O})$\\
    $TS \leftarrow updateArchive(P)$
}

\Return TS
\end{algorithm}

\subsection{Transforming Log Events to Traces}
As described in Section~\ref{sec:background_related}, a state machine is learned from traces of symbols. In the context of this work, each symbol in a trace represents an arbitrary log event outputted by the SUT, and each trace represents a sequential behavior exhibited by the SUT (based on the sequence of log events). We use a state machine to model these sequential behaviors within the SUT.

To transform event logs into traces, we first group the log events based on the start and end timestamps of the test cases. This is done to capture the sequential behavior exhibited by the SUT after running each test case. MISH requests the test cases' start and end timestamps from EvoMaster and temporarily stores them in memory (see line 3 in Algorithm~\ref{algo:mish}). We compare each log event's timestamp against each test case's start and end timestamp. If the log event occurred within this period, we add it to the group created for this test case. We keep track of the order in which the log events are added to the group so as not to disrupt the sequential behavior captured in these log events. Note that each test case is executed sequentially; therefore, an arbitrary log event (occurring at a given timestamp) cannot be added to two groups. 

Once the groups are formed, we proceed to generate the traces (line 4 in Algorithm~\ref{algo:mish}). We generate one trace for each test case and its corresponding log group. As each log event is a string containing an arbitrary number of characters, we need to transform it into a symbol for efficient learning of a state machine. We use the Drain algorithm proposed by He et al.~\cite{He_2017_Drain} to learn templates of log events and use these templates to map each log event to a single symbol. Specifically, we opted to use Drain 3 from LogPai~\cite{LogPai_Drain3} (version 0.9.11) for learning log templates as this version enables us to learn log templates on the fly and thus fits very well with the continuous learning of a state machine from log events. Each log statement in the group is mapped (in sequential order) to its corresponding log template symbol to form the trace produced by a test case. If the group for an arbitrary test case is empty, we use a single ``\textit{None}" word to create a trace of size one. This is done to make sure that the test case is not skipped and that a fitness value will still be assigned for this test case. The ``None" word is mapped to its symbol using Drain3. Timestamps collected for the test cases and groups of log statements are cleared from memory once we have finished generating the corresponding traces.

\subsection{Continuous Learning of State Machine}
MISH employs the streaming heuristic proposed by Baumgartner and Verwer~\cite{Baumgartner_2023_Learning} to continuously learn/update a state machine from the stream of traces produced by the SUT. For each generation of $N$ test cases, $N$ corresponding traces are generated and forwarded to the model inference framework, which updates the state machine incrementally (line 6 in Algorithm~\ref{algo:mish}). Once the state machine has been updated, the framework sends a response to MISH, which then uses the updated model to compute the fitness values for the test cases. This ensures that the state machine evolves dynamically alongside the test generation process. The fitness computation process is described in the next Subsection.

\subsection{Defining Fitness using State Machine}
Once the state machine has been updated, MISH forwards the traces (generated from the test cases) to the model inference framework and requests it to compute their fitness (line 7 in Algorithm~\ref{algo:mish}). The fitness value computed for a particular trace is assigned as the fitness value for the corresponding test case. Fitness is determined based on the path (the sequence of states) traversed by the trace within the state machine. The primary objective of MISH is to generate system-level test cases that trigger deep and interconnected interactions between microservices. Each state in the state machine represents a unique state of the SUT, and its visit frequency is computed from the traces used to update the model. We assume that test cases that lead the system into less frequently visited states (or traverse paths containing more of such states) are more likely to reveal complex and interconnected behaviors, thus contributing to higher system-level coverage. Consequently, MISH assigns higher fitness values to test cases that visit (or traverse paths through) infrequent states.

There are several ways to compute the fitness of a trace (and thus a test case) based on the state visit frequencies. We implemented two distinct fitness functions within the model inference framework. Below, we briefly describe each fitness function and provide an example of how the fitness value is calculated for an arbitrary trace using the corresponding function. For the example, we refer to the state machine and trace illustrated in Figure~\ref{fig:example_sm_proxyprint}. Note that all traces pass through the start state (e.g., state 0); therefore, this state is excluded from fitness computations.


\textbf{\textit{Lower Than Median (LM)}}: This fitness function first computes the median state visit frequency for the path taken by an arbitrary trace $t$ in the state machine. Then, it computes how many states in the path have a lower visit frequency than the median. 
To mitigate the tendency of MISH to favor longer traces, this count is divided by the total number of states visited by the trace. 
If the trace only visits one state, its fitness value is the inverse of its state visit frequency, mitigating the potential bias toward short traces that visit a single, highly frequent state.
Using the example state machine and trace $t$ from Figure~\ref{fig:example_sm_proxyprint}, the states visited by $t$ are 11, 12, 13, and 11. Their visit frequencies are 15, 15, 6, and 15, respectively. The median visit frequency for this trace is 15. Only one state (13) has a lower visit frequency than 15. The fitness of trace $t$ in this example is then $\frac{1}{4}{=}0.25$.

\textbf{\textit{Weighted State Visit Frequencies (WS)}}: this function first collects the visit frequency of the states visited by an arbitrary trace $t$. These frequencies are sorted in ascending order. A weighted sum is then calculated by multiplying each state's visit frequency by its index in the sorted list (starting from 1). The fitness value of the trace is defined as the inverse of this weighted sum.
Using the same trace $t$ from Figure~\ref{fig:example_sm_proxyprint}, the sorted state visit frequencies are 6, 15, 15, and 15. The weighted sum of these frequencies is then $(6\times1) + (15\times2) + (15\times3) + (15\times4)=141$. The fitness of $t$ is then $\frac{1}{141}{\approx}0.007$.



\subsection{Evolving and Selecting Test Cases}
In each iteration, MISH generates an offspring population based on the current population of test cases (see line 9 in Algorithm~\ref{algo:mish}). The offspring population has the same size as the current population. Specifically, MISH uses tournament selection to select candidate test cases for the offspring population. MISH then creates a copy of the selected candidate and introduces a mutation in the copied test case. The copied test case is then added to the offspring population. MISH selects 90\% of the offspring through tournament selection, and the remainder of the offspring population is filled by sampling test cases to create more diverse individuals for better exploration. We employ the same sampling strategy as MOSA. MISH then computes the fitness of the offspring population following the procedure described previously. Finally, MISH selects the new population (see line 16 of in Algorithm~\ref{algo:mish}) of test cases for the next iteration by taking the top $S$ test cases from the union of the current population and the offspring population, where $S$ denotes the population size. 


\section{Empirical Study}
This section details the empirical study we carried out to evaluate MISH's feasibility and effectiveness for generating system-level test cases. We first outline the research questions addressed in the study. Next, we detail the selection of applications and parameter settings used in EvoMaster for evaluation. Finally, we describe the metrics employed to compare the performance of MISH against MOSA.

\subsection{Research Questions}
The study aims to answer the following research questions:
\begin{itemize}
    \item \textbf{RQ1}: \textit{How effective is MISH in generating system-level test cases compared to the state-of-the-art MOSA algorithm?}
    \item \textbf{RQ2}: \textit{In which scenarios does MISH demonstrate a good/poor performance?}
    \item \textbf{RQ3}: \textit{How does MISH's coverage (test targets) over time compare to that of the state-of-the-art MOSA algorithm?}
\end{itemize}

\textbf{RQ1} focuses on evaluating the overall effectiveness of MISH in generating test cases that achieve high system-level coverage and reveal complex behaviors in REST APIs compared to MOSA. With \textbf{RQ2}, we aim to investigate in detail the scenarios in which MISH generates effective test cases and the scenarios in which MISH struggles with generating effective test cases. Finally, \textbf{RQ3} aims to evaluate and compare the efficiency of MISH in achieving test coverage over time, in comparison to MOSA.

\subsection{Selection of Applications}
To evaluate MISH's effectiveness in generating system-level test cases, we benchmark MISH on six real-world microservice applications from the EMB dataset. This dataset is created by the same authors as EvoMaster and was specifically designed to benchmark test case generation approaches of EvoMaster~\cite{Arcuri_2023_EMB}. The version number of the dataset is synced with the version number of EvoMaster. As MISH is developed within EvoMaster, we have selected this dataset for our evaluation. Specifically, we used version 2.0 of the dataset as this was the latest stable release at the time of our implementation. While version 3.2 of EvoMaster was released at the time of writing this work, we found it does not report a specific category of faults (Potential Faults) that we use to evaluate MISH's effectiveness (see Subsection~\ref{subsec:comparison_metrics} in the description of the metrics). Additionally, we encountered errors when attempting to write test cases to file. These two points led us to use an earlier version of the dataset for evaluation instead of the latest version. We selected only the applications in which REST APIs are used, as MISH is designed to test REST APIs. The six real-world applications that we have selected are:
\begin{itemize}
    \item \textit{CatWatch}: a web application that collects, processes, and stores statistics for a GitHub organization. The statistics can be requested via the REST API.
    \item \textit{Features-Service}: a microservice application for managing product Feature Models. The REST API is used to define and query products and product configurations.
    \item \textit{Scout-API}: A microservice application that provides a REST API to interact with a web service named "Scout"
    \item \textit{ProxyPrint}: a web service for comparing and making requests to different print shops
    \item \textit{LanguageTool}: an application that uses a REST API to do style and grammar checks for a piece of text.
    \item \textit{OCVN}: a dashboard for conducting visual analytics on public procurement data of Vietnam. 
\end{itemize}

All applications are built with JDK 8 following the instructions provided on the repository of the EMB dataset~\cite{Arcuri_EMB_repo}. Table~\ref{table:sut_stats} depicts the recent statistics collected for the selected applications from the EMB dataset~\cite{Arcuri_2024_Advanced}.

\begin{table}[t]
    \caption{Recent statistics collected by Arcuri et al. for the selected applications from the EMB dataset.}
    \label{table:sut_stats}
    \centering
    \resizebox{\columnwidth}{!}{
        \begin{tabular}{@{}lccc@{}}
        \toprule
            Application              & \# Source Files & \# LOCs & \# API Endpoints \\ \midrule
            CatWatch         & 106            & 9636   & 14                 \\
            Features-Service & 39             & 2275   & 18                 \\
            Scout-API        & 93             & 9736   & 49                 \\
            ProxyPrint       & 73             & 8338   & 74                 \\
            LanguageTool     & 1385           & 174781 & 2                  \\
            OCVN             & 526            & 45521  & 258                \\ \bottomrule
        \end{tabular}
    }
\end{table}

\subsection{Version \& Parameter Settings for EvoMaster}
We used EvoMaster version 2.0 to implement and benchmark MISH for the reason provided in the previous subsection.
We ran EvoMaster using mostly the default parameter settings, with a few modifications to the following parameters: search budget, test-case minimization, tournament selection size, and some other options that are only turned on by default for the MIO algorithm. The default search budget of EvoMaster is set to one minute. We changed the search budget to 60 minutes to give each EA enough time to initialize and explore the search space. This extended search budget aligns with previous studies~\cite{Arcuri_2024_Advanced, Golmohamaddi_2024_On_the_Impact,Seran_2024_Search-Based}. We opted to use time as the criterion for the search budget as we deem this to be a fairer comparison between different EAs. 
Alternative stopping criteria include the number of HTTP calls made by the algorithm, as used by Marculescu et al.\cite{Marculescu_2022_On_the_fault}. However, MISH would exhaust such a budget quickly because it uses the sequential patterns of REST API calls (captured in the state machine) to define the fitness of a test case. This rapid exhaustion would limit MISH's ability to test deeper sequences of interactions between microservices. Another stopping criterion is the number of iterations or generations an EA has executed\cite{Ravber_2022_Maximum}. 
However, this approach does not consider the complexity of the system under test; some systems require more generations to trigger deep and interesting behaviors. 
Using the number of generations might, therefore, halt the exploration prematurely for complex systems. Moreover, practitioners often allocate a fixed amount of time to run an EA, as argued by Stallenberg et al.~\cite{Stallenberg_2021_Improving}, making time a practical and equitable stopping criterion.

\textit{Test case minimization}.
We turned off the post-process minimization process of EvoMaster. This process splits larger test cases with multiple HTTP calls into smaller test cases with one single HTTP call at the end of the search. This type of splitting does exactly the opposite of what we are trying to tackle with MISH: it breaks the sequential patterns that MISH tries to capture in the REST API calls. Additionally, we observed that enabling test-case minimization led to a reduction in the coverage achieved by both MISH and MOSA compared to when minimization was disabled.

\textit{Tournament selection}.
We modified the default tournament selection size from ten to four for our experiments. A tournament size of ten is considered too large for a single-objective EA like MISH, potentially hindering its ability to effectively select and evolve promising test cases~\cite{lavinas2018experimental}.

\textit{Configuration options}.
We activated several non-default configuration options in EvoMaster, such as smarter sampling of individuals, enhanced mutation operators, and optimized database queries. We found that activating these options led to consistently better results (i.e., higher coverage) for both MISH and MOSA in our preliminary experiments.



\subsection{Execution Procedure of Each Algorithm}
To answer our research questions, we ran each algorithm (MOSA and MISH) 20 times on each selected microservice application to mitigate the effects of stochastic operations (e.g., mutation) on performance results. Each run is executed independently, i.e., the application and the EA are reset after each run to make sure that they are both started in a clean state for collecting performance statistics. Since MISH implements two distinct fitness functions, each version of MISH was also run 20 times per application. In total, we executed 360 runs, which would take approximately 15 days to complete if run sequentially.
To speed up the evaluation process, we created a Docker image that enables us to parallelize the runs. This was inspired by the experimental setup used by Arcuri et al.~\cite{Arcuri_2024_Advanced}. Each Docker container executes 20 runs of an algorithm on a particular application. We make our Docker image and source code publicly available\cite{Cao_2024_MISH_docker,Cao_2024_MISH_GitHub} to promote the reproducibility of our study. Our Docker image does not include the option to run OCVN, as it spawns a Docker container. We chose not to run Docker containers within another Docker container due to potential performance and compatibility issues. For experiments involving OCVN, we executed all three algorithms directly on the host machine, ensuring a fair comparison by using the same hardware specifications for all runs.

\subsection{Comparison Metrics}\label{subsec:comparison_metrics}
We collect and compare the algorithms based on the performance metrics used in the literature~\cite{Marculescu_2022_On_the_fault, Stallenberg_2021_Improving, Arcuri_2024_Advanced,Golmohamaddi_2024_On_the_Impact}:
\begin{itemize}
    \item Number of covered test targets: aggregated value denoting the (unique) number of search objectives (e.g. line, branch, faults, etc.) covered during the run.
    \item Number of potential faults: aggregated value denoting the (unique) number of 500 errors detected from the HTTP responses, and the (unique) number of discrepancies detected between API responses and the specifications in the OpenAPI schema.
    \item Number of potential 500 faults: the (unique) number of 500 error codes detected from HTTP responses.
\end{itemize}

To determine whether the results between the algorithms are significantly different, we performed unpaired Wilcoxon rank-sum tests with a threshold of 0.05 to compute the $\rho$-values. This is a non-parametric pairwise test for determining whether a pair of data distributions are significantly different from each other~\cite{Conover_1971_Practical}. Since MOSA is our baseline method and we have two different versions of MISH, we performed a pairwise comparison between each version of MISH and MOSA. Furthermore, to determine the probability that a sample of results from one method will exceed those from the other method, we computed the effect size of the result using the Vargha-Delaney statistic~\cite{Vargha_2000_A_Critique}


\section{Evaluation Results}
We now answer our research questions based on the results achieved from our empirical study. Each research question is answered separately in the following subsections.

\begin{table}[t]
\centering
\caption{Statistics are collected from the model inference framework.}
    \label{table:mish_training_statistics}
    \resizebox{\columnwidth}{!}{
        \begin{tabular}{lcc|cc|cc}
        \toprule
         \multirow{2}{*}{Application}     & \multicolumn{2}{c|}{Trace Length} & \multicolumn{2}{c|}{Learning Time (seconds)} & \multicolumn{2}{c}{Model Size} \\
              \cline{2-7}
              & Median           & IQR            & Median                & IQR                  & Median         & IQR           \\ \midrule
        CatWatch         & 5.00             & 7.00           & 0.0187                & 0.0039               & 29.00          & 13.50         \\
        Features-Service & 2.00             & 2.00           & 0.0031                & 0.0005               & 22.00          & 3.00          \\
        Scout-API        & 10.00            & 4.00           & 0.0107                & 0.0025               & 20.00          & 8.00          \\
        ProxyPrint       & 3.00             & 8.00           & 0.0026                & 0.0007               & 44.00          & 7.00          \\
        LanguageTool     & 1.00             & 0.00           & 0.0370                & 0.0353               & 93.00          & 99.00         \\
        OCVN             & 1.00             & 0.00           & 0.0046                & 0.0006               & 8.00           & 4.00          \\ \bottomrule
        \end{tabular}
    }
\end{table}
 
\begin{table}[t]
    \centering
    \caption{Median number of covered targets.}
    \label{table:median_targets}
    \resizebox{\columnwidth}{!}{
        \begin{tabular}{@{}lcc|cc|cc@{}}
        \toprule
        \multirow{2}{*}{Application}                 & \multicolumn{2}{c|}{MOSA} & \multicolumn{2}{c|}{MISH - LM} & \multicolumn{2}{c}{MISH - WS} \\
        \cline{2-7}
        & Median      & IQR         & Median         & IQR           & Median        & IQR           \\ \midrule
        CatWatch         & \textbf{2014.0}      & 471.25      & 1703.5         & 51.25         & 1725.5        & 85.5          \\
        Features-Service & 879.0       & 6.25        & 883.0          & 4.75          & \textbf{885.0}         & 1.5           \\
        Scout-API        & \textbf{2223.0}      & 68.25       & 2073.0         & 85.0          & 2089.5        & 84.5          \\
        ProxyPrint       & 2922.5      & 308.5       & 3302.0         & 127.75        & \textbf{3377.5}        & 81.0          \\
        LanguageTool     & \textbf{3106.0}      & 10262.5     & 2927.0         & 86.0          & 2948.5        & 87.75         \\
        OCVN             & 4464.5      & 170.0       & \textbf{6358.0}         & 427.75        & 6083.5        & 230.75        \\ \bottomrule
        \end{tabular}
    }
\end{table}

\begin{table}[t]
    \centering
    \caption{Median number of potential faults.}
    \label{table:median_faults}
    \resizebox{\columnwidth}{!}{
        \begin{tabular}{@{}lcc|cc|cc@{}}
        \toprule
        \multirow{2}{*}{Application}                  & \multicolumn{2}{c|}{MOSA} & \multicolumn{2}{c|}{MISH - LM} & \multicolumn{2}{c}{MISH - WS} \\
        \cline{2-7}
        & Median       & IQR        & Median          & IQR          & Median         & IQR          \\ \midrule
        CatWatch         & \textbf{25.0}         & 1.25       & 24.0            & 2.0          & 24.0           & 1.0          \\
        Features-Service & \textbf{32.0}         & 0.0        & 31.0            & 0.5          & 31.5           & 1.0          \\
        Scout-API        & \textbf{79.0}         & 9.5        & 66.0            & 2.0          & 69.0           & 4.25         \\
        ProxyPrint       & 66.0         & 10.25      & 68.0            & 3.0          & \textbf{69.0}           & 2.25         \\
        LanguageTool     & \textbf{7.0}          & 2.0        & 6.5             & 1.0          & 6.0            & 0.0          \\
        OCVN             & 274.5        & 7.0        & \textbf{311.5}           & 13.0         & 303.0          & 10.5         \\ \bottomrule
        \end{tabular}
    }
\end{table}

\begin{table}[t]
    \centering
    \caption{Median number of potential 500 faults}
    \label{table:median_500_faults}
        \resizebox{\columnwidth}{!}{
        \begin{tabular}{@{}lcc|cc|cc@{}}
        \toprule
        \multirow{2}{*}{Application}              & \multicolumn{2}{c|}{MOSA} & \multicolumn{2}{c|}{MISH - LM} & \multicolumn{2}{c}{MISH - WS} \\
        \cline{2-7}
        & Median       & IQR        & Median          & IQR          & Median         & IQR          \\ \midrule
        CatWatch         & 10.0         & 1.25       & 10.0            & 2.0          & 10.0           & 1.25         \\
        Features-Service & 25.0         & 0.0        & 25.0            & 1.0          & 25.0           & 1.0          \\
        Scout-API        & \textbf{59.5}         & 9.0        & 47.0            & 2.25         & 50.0           & 4.25         \\
        ProxyPrint       & 32.0         & 5.25       & 33.0            & 2.0          & \textbf{34.0}           & 2.0          \\
        LanguageTool     & \textbf{5.0}          & 1.0        & 4.5             & 1.0          & 4.0            & 0.0          \\
        OCVN             & 49.0         & 4.5        & \textbf{86.5}            & 9.75         & 79.5           & 10.75        \\ \bottomrule
        \end{tabular}
    }
\end{table}

\subsection{RQ1: MISH's Performance In Generating Test Cases}
Table~\ref{table:mish_training_statistics} presents various learning statistics gathered from the model inference framework for each application. These represent the averages (medians and interquartile ranges) for one run on the respective applications. We observe that the time MISH requires to learn or update the state machine is in the order of milliseconds, which is negligible compared to the overall search budget of 60 minutes. This demonstrates the feasibility of integrating real-time state machine learning into an EA for system-level test case generation.


Tables~\ref{table:median_targets},~\ref{table:median_faults}, and~\ref{table:median_500_faults} present the performance of each algorithm for every application according to the respective comparison metric. The best-performing algorithm for each corresponding application is highlighted in bold. To address the outliers caused by random processes in EAs, we report the medians along with their corresponding interquartile ranges (IQR) instead of the mean.


From Table~\ref{table:median_targets}, both versions of MISH achieve higher coverage than MOSA on Features-Service, ProxyPrint, and OCVN. On the remaining applications, MISH demonstrates coverage results that are competitive with MOSA. MISH achieved, on average, a minimal increase (0.68\%) in coverage on Feature-Service when compared to MOSA. However, MISH achieved a much larger increase in coverage on ProxyPrint and OCVN (15.7\% and 42.4\%, respectively). Regarding fault detection capabilities (Tables~\ref{table:median_faults} and~\ref{table:median_500_faults}), MISH again outperforms MOSA on ProxyPrint and OCVN, while detecting a comparable number of faults as MOSA on the other applications. Specifically, for the potential faults detected on ProxyPrint and OCVN, MISH detected (on average) 4.5\% and 13.5\% more faults than MOSA, respectively. For the potential 500 faults, MISH detected (on average) 6.3\% more faults on ProxyPrint and 76.5\% more faults on OCVN. 
Tables~\ref{table:wilcoxon_test_targets},~\ref{table:wilcoxon_test_faults}, and~\ref{table:median_500_faults} present the results of the statistical tests conducted for each comparison metric. Significant $\rho$-values ($<0.05$) are highlighted in gray. As expected, there is no statistical significance in MISH's coverage performance on Feature-Service as it achieved only a minimal increase in coverage. However, MISH performed significantly better on ProxyPrint and OCVN across all three metrics, each associated with a \textit{large} effect size. The results of our statistical tests confirm our findings presented in Tables~\ref{table:median_targets},~\ref{table:median_faults} and~\ref{table:median_500_faults}.

Although MISH achieved lower coverage in the latter applications, it is noteworthy that MISH can still detect a similar number of faults in these instances. The lower coverage performance of MISH is expected as it is a single-objective EA and single-objective EAs typically get stuck very quickly in local optima. In contrast, many-objective EAs (e.g., MOSA) can use multiple criteria to generate a wider diversity of individuals, resulting in better exploration~\cite{Black_2004_Bi-Criteria,Soltani_2018_Single-Objective,Yoo_2007_Pareto}.

\begin{table}[t]
    \caption{Statistical significance (($\rho$-values and $\hat{A}_{12}$-values)) for the number of covered targets.}
    \label{table:wilcoxon_test_targets}
\centering
    \resizebox{\columnwidth}{!}{
        \begin{tabular}{@{}lcc|cc@{}}
        \toprule
        \multirow{2}{*}{Application}                 & \multicolumn{2}{l|}{MISH - LM vs. MOSA}       & \multicolumn{2}{l}{MISH -WS vs. MOSA}         \\
        \cline{2-5}
        & $\rho$-value                 & $\hat{A}_{12}$ & $\rho$-value                 & $\hat{A}_{12}$ \\ \midrule
        CatWatch         & \multicolumn{1}{c|}{\cellcolor{gray!55}$<0.01$} & 0.21 (large)   & \multicolumn{1}{c|}{\cellcolor{gray!55}$<0.01$} & 0.27 (medium)  \\
        Features-Service & \multicolumn{1}{c|}{0.47}    & 0.62 (small)   & \multicolumn{1}{c|}{0.47}    & 0.87 (large)   \\
        Scout-API        & \multicolumn{1}{c|}{\cellcolor{gray!55}$<0.01$} & 0.01 (large)   & \multicolumn{1}{c|}{\cellcolor{gray!55}$<0.01$} & 0.10 (large)   \\
        ProxyPrint       & \multicolumn{1}{c|}{\cellcolor{gray!55}$<0.01$} & 0.89 (large)   & \multicolumn{1}{c|}{\cellcolor{gray!55}$<0.01$} & 0.95 (large)   \\
        LanguageTool     & \multicolumn{1}{c|}{\cellcolor{gray!55}$<0.01$} & 0.10 (large)   & \multicolumn{1}{c|}{\cellcolor{gray!55}$<0.01$} & 0.13 (large)   \\
        OCVN             & \multicolumn{1}{c|}{\cellcolor{gray!55}$<0.01$} & 1.00 (large)   & \multicolumn{1}{c|}{\cellcolor{gray!55}$<0.01$} & 1.00 (large)   \\ \bottomrule
        \end{tabular}
    }
\end{table}

\begin{table}[t]
\centering
    \caption{Statistical significance ($\rho$-values and $\hat{A}_{12}$-values) computed for the number of potential faults.}
    \label{table:wilcoxon_test_faults}
    \resizebox{\columnwidth}{!}{
        \begin{tabular}{lcc|cc}
        \hline
         \multirow{2}{*}{Application}   & \multicolumn{2}{l|}{MISH - LM vs. MOSA}     & \multicolumn{2}{l}{MISH -WS vs. MOSA}      \\
         \cline{2-5}
                       & $\rho$-value              & $\hat{A}_{12}$ & $\rho$-value              & $\hat{A}_{12}$ \\ \hline
        CatWatch         & \multicolumn{1}{c|}{0.06} & 0.37 (small)   & \multicolumn{1}{c|}{0.06} & 0.37 (small)   \\
        Features-Service & \multicolumn{1}{c|}{\cellcolor{gray!55}0.01} & 0.23 (large)   & \multicolumn{1}{c|}{\cellcolor{gray!55}0.01} & 0.40 (small)   \\
        Scout-API        & \multicolumn{1}{c|}{\cellcolor{gray!55}$<0.01$} & 0.01 (large)   & \multicolumn{1}{c|}{\cellcolor{gray!55}$<0.01$} & 0.02 (large)   \\
        ProxyPrint       & \multicolumn{1}{c|}{\cellcolor{gray!55}0.02} & 0.69 (medium)  & \multicolumn{1}{c|}{\cellcolor{gray!55}0.02} & 0.81 (large)   \\
        LanguageTool     & \multicolumn{1}{c|}{\cellcolor{gray!55}0.01} & 0.29 (medium)  & \multicolumn{1}{c|}{\cellcolor{gray!55}0.01} & 0.18 (large)   \\ 
        OCVN             & \multicolumn{1}{c|}{\cellcolor{gray!55}$<0.01$} & 1.00 (large)   & \multicolumn{1}{c|}{\cellcolor{gray!55}$<0.01$} & 1.00 (large)   \\ \bottomrule
        \end{tabular}
    }
\end{table}

\begin{table}[t]
\centering
    \caption{Statistical significance ($\rho$-values and $\hat{A}_{12}$-values) computed for the number of potential 500 faults.}
    \label{table:wilcoxon_test_500_faults}
    \resizebox{\columnwidth}{!}{
        \begin{tabular}{lcc|cc}
        \hline
        \multirow{2}{*}{Application}      & \multicolumn{2}{l|}{MISH - LM vs. MOSA}        & \multicolumn{2}{l}{MISH -WS vs. MOSA}      \\
        \cline{2-5}
        & $\rho$-value              & $\hat{A}_{12}$    & $\rho$-value              & $\hat{A}_{12}$ \\ \hline
        CatWatch         & \multicolumn{1}{c|}{0.35} & 0.45 (negligible) & \multicolumn{1}{c|}{0.35} & 0.41 (small)   \\
        Features-Service & \multicolumn{1}{c|}{0.36} & 0.44 (negligible) & \multicolumn{1}{c|}{0.36} & 0.63 (small)   \\
        Scout-API        & \multicolumn{1}{c|}{\cellcolor{gray!55}$<0.01$} & 0.01 (large)      & \multicolumn{1}{c|}{\cellcolor{gray!55}$<0.01$} & 0.02 (large)   \\
        ProxyPrint       & \multicolumn{1}{c|}{\cellcolor{gray!55}0.03} & 0.66 (small)      & \multicolumn{1}{c|}{\cellcolor{gray!55}0.03} & 0.81 (large)   \\
        LanguageTool     & \multicolumn{1}{c|}{0.18} & 0.41 (small)      & \multicolumn{1}{c|}{0.18} & 0.25 (large)   \\
        OCVN             & \multicolumn{1}{c|}{\cellcolor{gray!55}$<0.01$} & 1.00 (large)   & \multicolumn{1}{c|}{\cellcolor{gray!55}$<0.01$} & 1.00 (large)   \\ \bottomrule
        \end{tabular}
    }
\end{table}

\subsection{RQ2: Where does MISH exhibit good/poor performance?}
We analyzed scenarios where MISH performs well or poorly by examining one application for each case in detail. For strong performance, we selected OCVN, while for poor performance, we analyzed Scout-API. These applications were chosen due to the notable (significant) differences in the number of faults identified by MISH and MOSA.

\subsubsection*{\textbf{Perfomance on OCVN}}
\begin{figure}[t]
    \centering
    \includegraphics[width=0.39\linewidth]{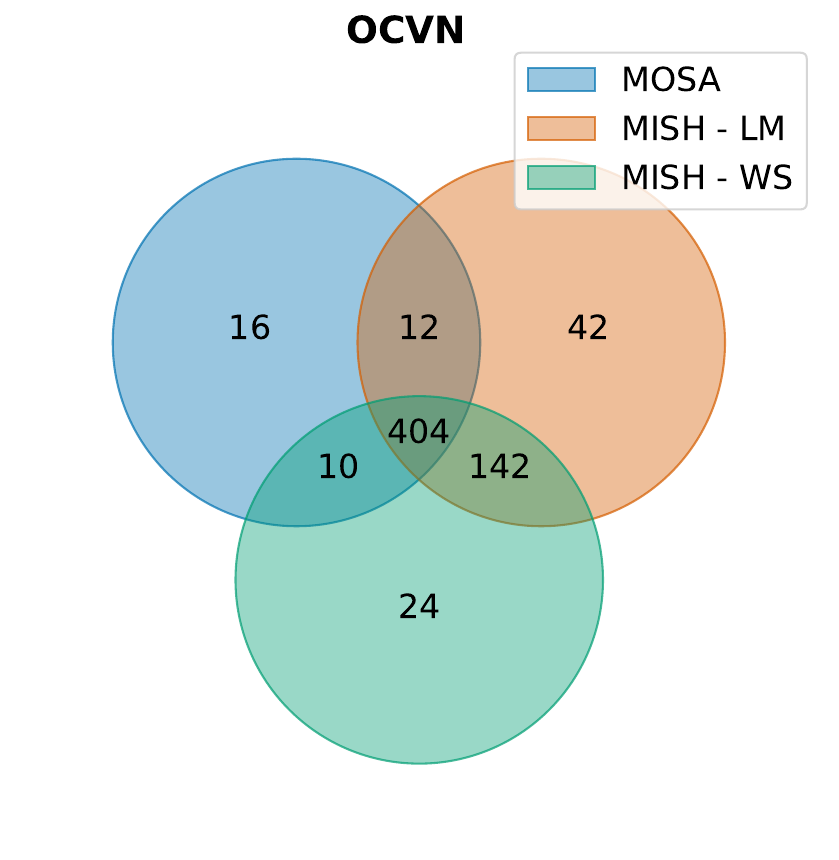}
    \caption{Venn diagram for the faults detected by each algorithm on OCVN.}
    \label{fig:ocvn_faults_venn_diagram}
\end{figure}

To understand MISH's superior performance on OCVN, we compared the faults identified by MISH-LM and MISH-WS with those identified by MOSA. Specifically, we analyzed the unique set of fault IDs detected across all runs of this application. 
Figure~\ref{fig:ocvn_faults_venn_diagram} presents the Venn diagram on the faults detected by each algorithm in OCVN. As we can observe, each version of MISH has identified more unique faults than MOSA. Notably, MISH-LM identified more than twice as many faults as MOSA. 

Upon analyzing the faults identified by MOSA, we found they were related to seven distinct fault locations in the code. Of these, only two fault locations were associated with test cases involving multiple (and sequential) REST API calls. In contrast, the faults identified by MISH were spread across 38 distinct fault locations. A deeper analysis of these faults revealed a common starting pattern in the REST API calls: the first call logs in as an administrator, saving authentication information in a cookie that is subsequently used for the remaining REST API calls. This indicates a dependency between the initial call and the subsequent calls in the tests.

MISH's design explicitly aims to capture such dependency patterns in REST API calls to trigger interconnected behaviors between microservices. By leveraging model inference to identify and utilize these patterns, MISH was able to generate more effective system-level test cases. Notably, no similar patterns were observed in the test cases associated with faults identified solely by MOSA, highlighting the strength of MISH's approach.

An important observation from our evaluation is that MISH performs particularly well on applications with numerous API endpoints, such as ProxyPrint and OCVN. Additional investigation into the faults identified only by MISH on ProxyPrint revealed a pattern similar to OCVN: the first call logs in as an administrator, followed by a sequence of dependent API calls. We believe that MISH's superior performance on these two applications is due to its ability to capture and exploit the dependencies between REST API calls, which MOSA (and its many-objective design) does not explicitly account for.

\subsubsection*{\textbf{Performance on Scout-API}}
\begin{figure}[!t]
    \centering
    \includegraphics[width=0.39\linewidth]{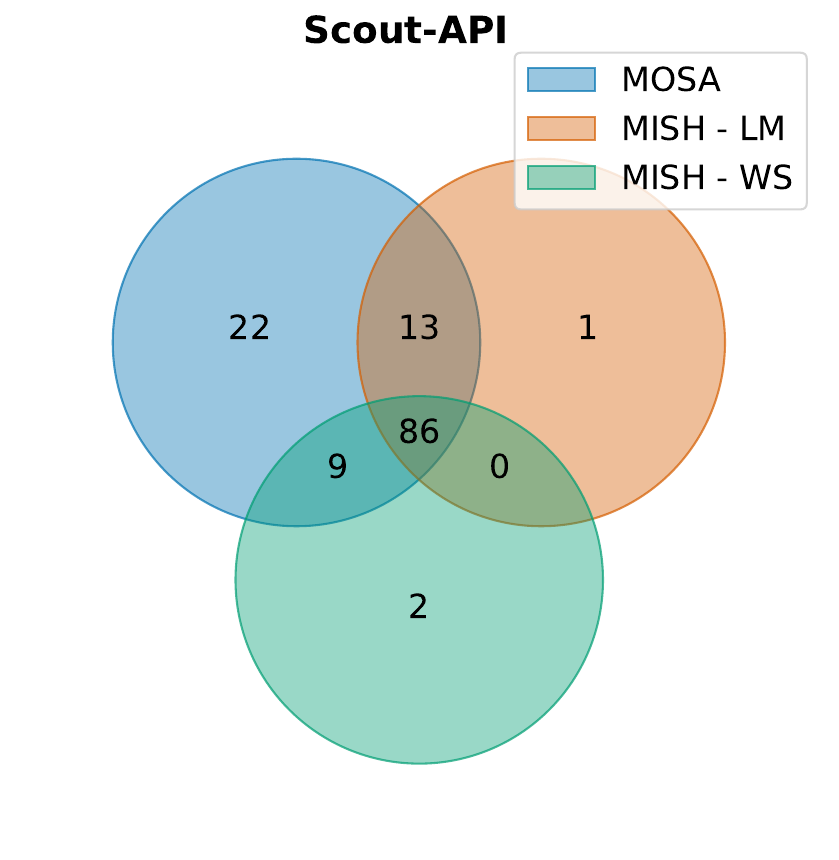}
    \caption{Venn diagram for the faults detected by each algorithm on Scout-API.}
    \label{fig:scout-api_faults_venn_diagram}
\end{figure}

To understand MISH's poor performance on Scout-API, we applied the same methodology used for the analysis of OCVN. Figure~\ref{fig:scout-api_faults_venn_diagram} presents a Venn diagram showing the distribution of fault IDs identified by the three algorithms.
From the diagram, we observe that 22 unique fault IDs were identified solely by MOSA, whereas both versions of MISH together only identified three unique fault IDs that MOSA did not. Among the 22 unique faults found by MOSA, all are potential 500 faults, and 17 of these faults are related to the location shown in Listing~\ref{lst:scout-api_common_fault_location_mosa}. At this location, MOSA executed different actions in the REST API calls to trigger these faults.

For the three unique faults identified by MISH, only MISH-LM managed to find a potential 500 fault at the same location as MOSA. However, MISH-LM executed a different action to trigger it. This suggests that MOSA's superior performance can be attributed to its many-objective nature and access to additional information, such as source code. With source code access, MOSA can compute fine-grained metrics like branch distances, which can guide test-case generation more effectively by enabling diverse actions in REST API calls. MOSA leveraged this capability to discover multiple faults at the same location by varying the actions in the API calls.

In contrast, MISH, as a single-objective EA, does not have access to such information. It relies solely on log statements and the state machine inferred from them to guide test-case generation. As a result, MISH lacks the ability to make fine-grained modifications to test cases, limiting its capacity to discover faults in complex scenarios like those in Scout-API.

\begin{lstlisting}[basicstyle=\ttfamily\footnotesize, caption={Common fault location in the 22 faults identified only by MOSA}, label={lst:scout-api_common_fault_location_mosa}, frame=single, captionpos=b, belowcaptionskip=3.5em]
devscout/scoutapi/dao/UserDao_30_readUserByIdentity
\end{lstlisting}

\textit{\textbf{Co-factor Analysis}}. To statistically understand whether there is a significant interaction between the search algorithms (MOSA, MISH-LM, or MISH-WS), the size of the application under tests (measures as the number of endpoints), and the achieved coverage, we conducted a two-way \textit{permutation test}~\cite{crawley2014statistics}. This test is a non-parametric alternative to the two-way ANOVA test and does not require the data to be normally distributed (unlike ANOVA). 
It enables us to determine whether the different algorithms achieve significantly different coverage and whether they behave statistically differently depending on the size of the application under test. For this test, we set the threshold $p$-value$\leq$0.05.

The test revealed that the algorithms have a significant impact on the achieved coverage ($p$-value=0.04), confirming the results of the Wilcoxon test. Besides, there is a significant relation between the coverage achieved by the different algorithms and the size of the applications under test ($p$-value$<$0.01). This is expected as it is more challenging to achieve higher coverage for larger applications. The two-way test also reveals that there is a significant interaction between the algorithms and the size of the applications (independent variable) on the achieved coverage (dependent variable), returning $p$-value$<$0.01. This confirms our observational results that MISH seems to perform better specifically for applications with a larger number of endpoints.

\subsection{RQ3: MISH's Coverage Over Time}
To evaluate MISH's efficiency in covering test targets, we analyzed the coverage data collected from all 360 runs of the experiments. In each run, EvoMaster reported the number of test targets covered by the algorithm and the corresponding search budget utilized. We grouped the coverage data by application and computed the average coverage over time for each algorithm on every application. Plotting the coverage over time enables us to assess how quickly each algorithm discovers previously unseen test targets.


\begin{figure}[t]
    \centering
    \includegraphics[width=0.78\columnwidth]{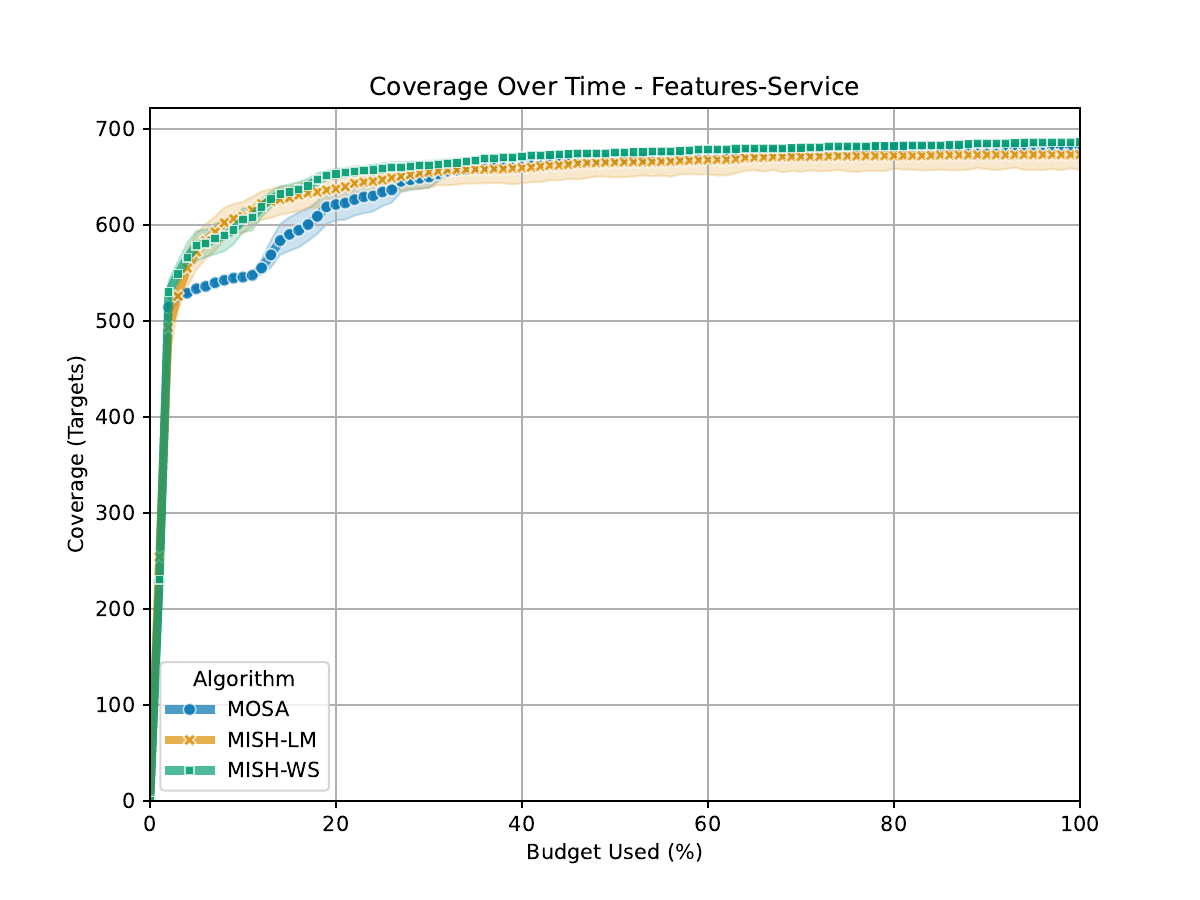}
    \caption{Average coverage over time on Features-Service}
    \label{fig:convergence_plot_features-service}
\end{figure}

\begin{figure}[t]
    \centering
    \includegraphics[width=0.78\columnwidth]{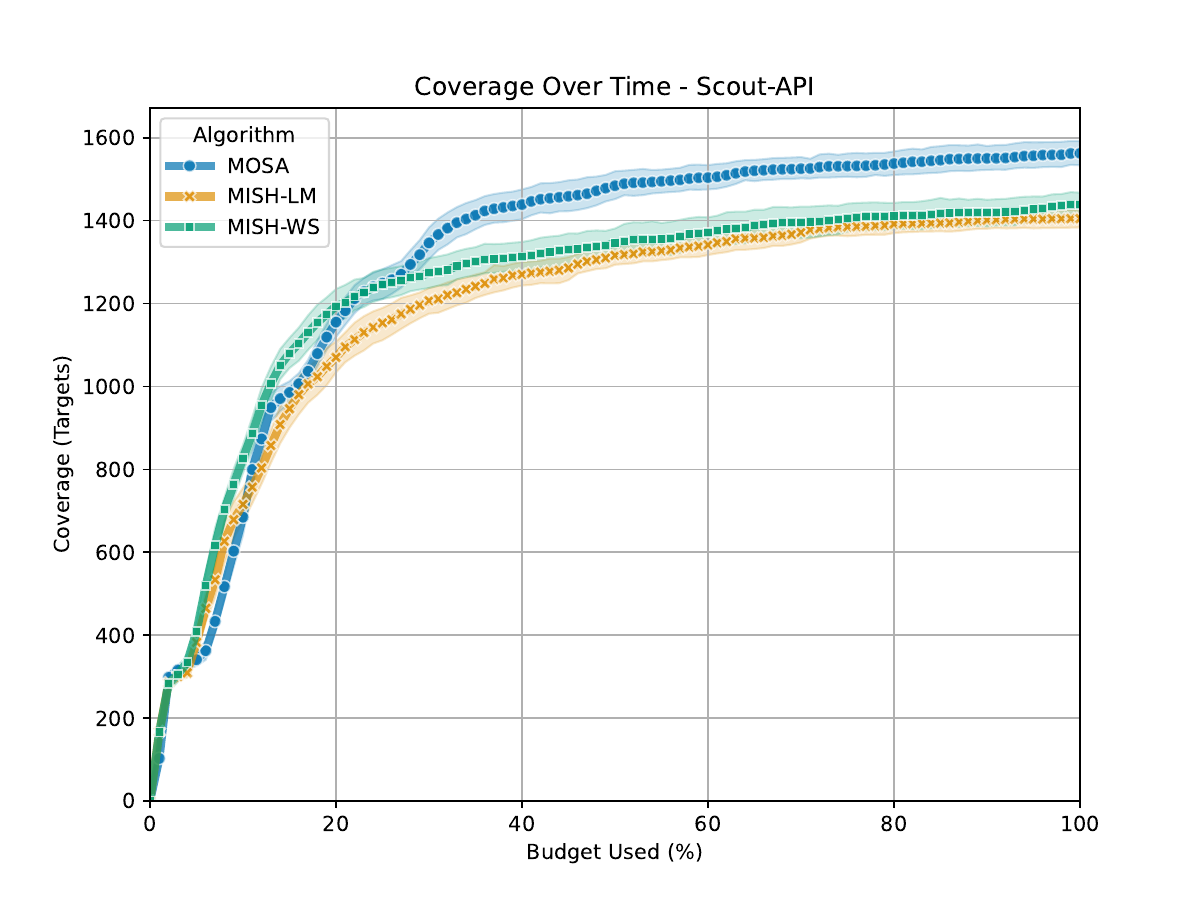}
    \caption{Average coverage over time on Scout-API}
    \label{fig:convergence_plot_scout-api}
\end{figure}

\begin{figure}[t]
    \centering
    \includegraphics[width=0.78\columnwidth]{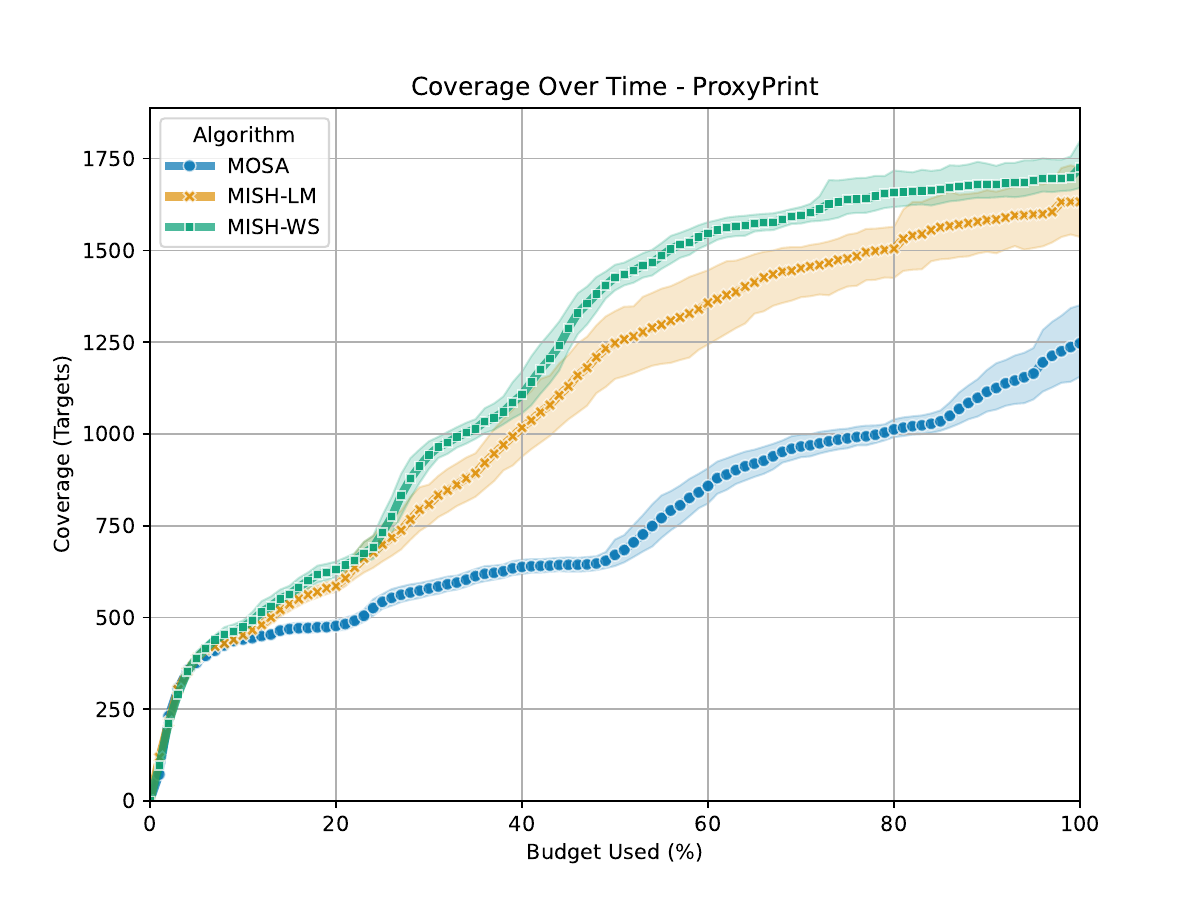}
    \caption{Average coverage over time on ProxyPrint}
    \label{fig:convergence_plot_proxyprint}
\end{figure}

\begin{figure}[t]
    \centering
    \includegraphics[width=0.78\columnwidth]{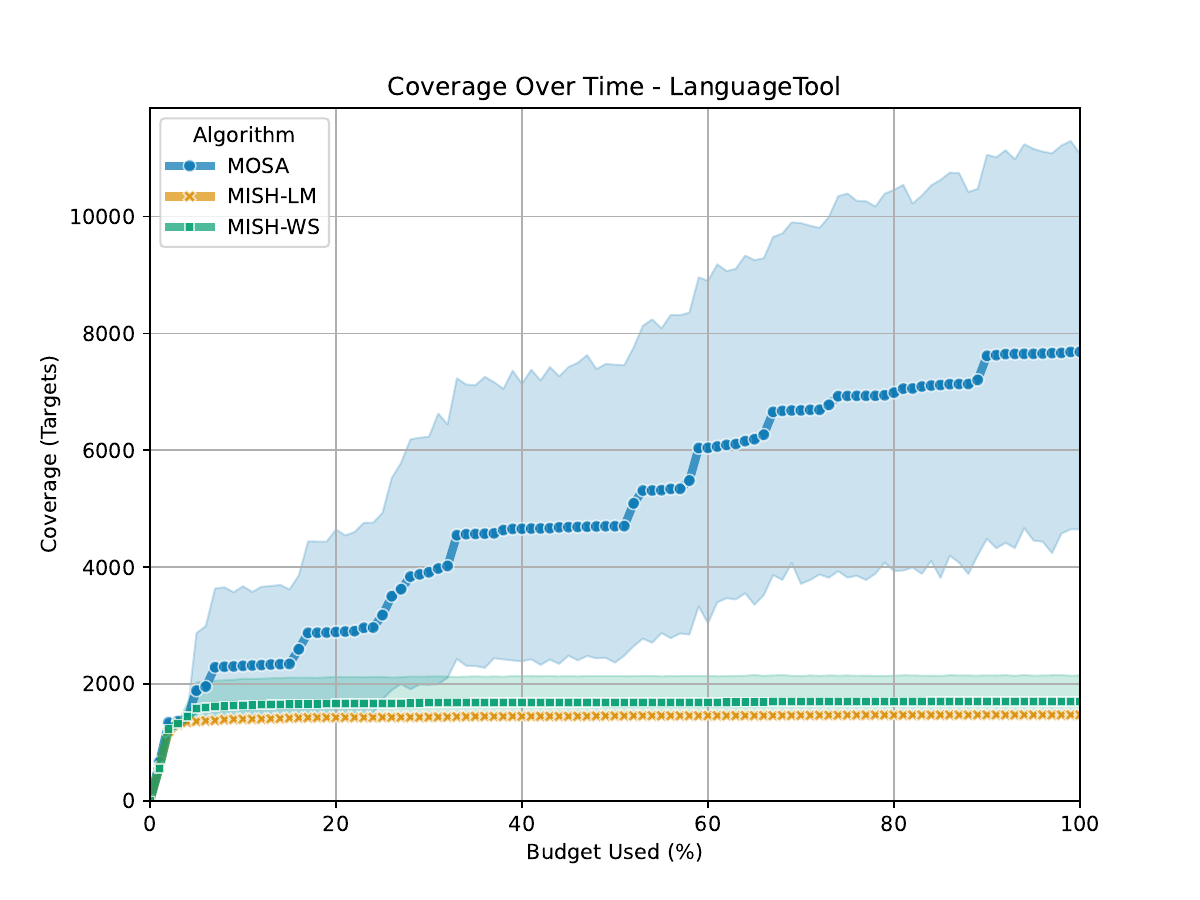}
    \caption{Average coverage over time on LanguageTool}
    \label{fig:convergence_plot_languagetool}
\end{figure}

\begin{figure}[t]
    \centering
    \includegraphics[width=0.78\columnwidth]{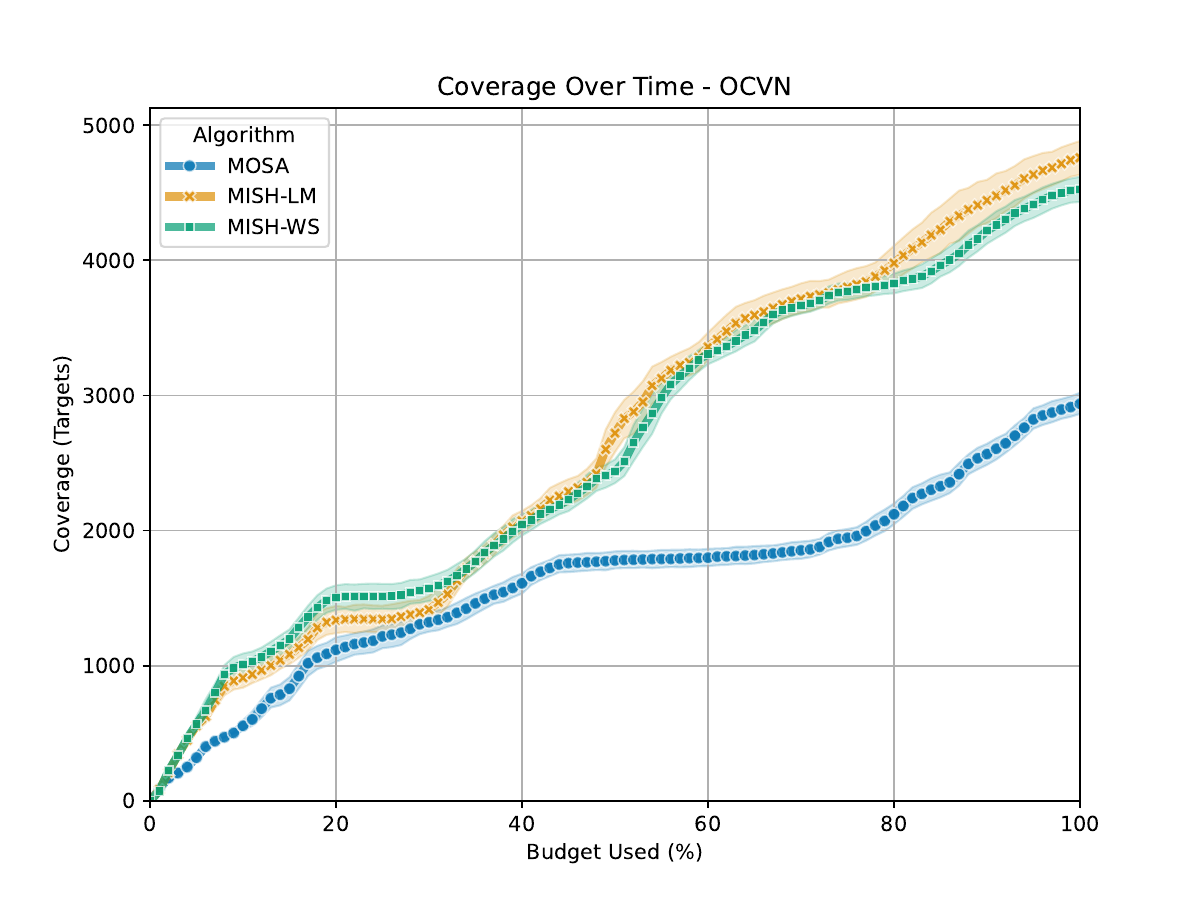}
    \caption{Average coverage over time on OCVN}
    \label{fig:convergence_plot_ocvn}
\end{figure}

Figures~\ref{fig:convergence_plot_features-service} to~\ref{fig:convergence_plot_ocvn} present the average coverage over time for all three algorithms on the respective applications. Confidence intervals are included to illustrate the variance in performance. These plots show that MISH discovers unseen test targets more quickly than MOSA on almost all applications. We have omitted the plot of CatWatch to conserve space.

The one exception is \textit{LanguageTool}, where MISH does not outperform MOSA in discovering unseen test targets. On this application, MISH's coverage performance reaches a plateau early in the run and remains stagnant for the rest of the execution. Two key factors contribute to MISH's performance on \textit{LanguageTool}. First, \textit{LanguageTool} has the largest number of source files and lines of code among the applications, which increases the complexity of the system under test. Second, \textit{LanguageTool} has only two API endpoints, limiting MISH's ability to generate diverse traces. MISH relies on log statements produced by these endpoints to create traces and update the state machine. Furthermore, MISH can only adjust the HTTP request parameters and actions directed at these endpoints to identify new test targets. This limitation hinders MISH's ability to uncover more varied behaviors, particularly when the API endpoints function independently and lack dependencies that might reveal sequential patterns.


In the worst-case scenario where the two endpoints are independent (i.e., one endpoint's behavior does not depend on the other), MISH is limited to testing each API endpoint in isolation. Due to the sheer size of \textit{LanguageTool}, MISH, as a single-objective evolutionary algorithm, is more likely to become trapped in a local optimum. Unlike MOSA, MISH lacks access to the source code, which MOSA uses to compute metrics such as branch distance to generate a more diverse set of test cases. This lack of diversity hinders MISH's ability to escape local optima, a limitation supported by the significant variance in MOSA's coverage performance on \textit{LanguageTool}.

For the other applications, the coverage trends show a consistent pattern: MISH initially discovers test targets faster than MOSA but eventually reaches a plateau for the remainder of the run. This behavior is expected given that MISH is a single-objective EA without access to additional information that could diversify the population of test cases. However, it is worth noting that MISH achieves its initial performance boost solely by leveraging log statements outputted by the application. This demonstrates MISH's ability to capture essential patterns early in the run, leading to early higher coverage.

These results suggest that MISH could benefit from a hybrid approach. A method that starts with MISH to capture essential patterns and then switches to MOSA to diversify the test cases and improve exploration could combine the strengths of both algorithms, potentially yielding better overall performance.

\section{Threat to Validity}

Concerning threats to \textit{construct validity}, we evaluated all evolutionary algorithms (EAs) using metrics widely employed in prior studies: the number of covered targets, the number of potential faults detected, and the number of potential 500 faults detected. For the stopping criterion, we utilized a commonly used approach based on a fixed search budget measured by runtime. Specifically, we set the search budget to 60 minutes per run, consistent with prior EvoMaster research.

A potential threat to \textit{internal validity} could arise from the integration of state machine learning into the EA for test-case generation. The additional learning time for the state machine could negatively impact the runtime of the EA. However, based on the average statistics collected from the model inference framework, the state machine learning time is in the order of milliseconds. Thus, we consider this impact to be minimal.

For threats to \textit{external validity}, we selected six applications from the well-established EvoMaster Benchmark (EMB), which has been extensively used in prior research. Our study focused on real-world applications, excluding any artificial or synthetic applications. The selected applications vary in size, including differences in the number of source files, lines of code, and API endpoints, which helps assess the generalizability of our algorithm across diverse scenarios.

To address threats to \textit{conclusion validity}, we followed best practices from prior studies on EvoMaster. Each algorithm was run 20 times per application, using different random seeds to account for the stochastic nature of EAs. For statistical analysis, we employed the Wilcoxon rank-sum test to evaluate the significance of differences between algorithms and the Vargha-Delaney effect size measure to quantify their magnitude. 
\section{Conclusion \& Future Work}
We presented MISH, a novel approach that uses real-time automaton learning as a search heuristic to guide the test case generation within EvoMaster. MISH continuously updates an automaton by gathering logs produced by the SUT after running a generation of test cases. MISH computes fitness for the test cases based on the path traversed by the log statements within the automaton. MISH uses the automaton to capture the sequence of REST API calls needed to trigger deep and connected behaviors between microservices. 

We implemented MISH within EvoMaster and empirically evaluated its effectiveness in generating test cases on six microservice applications from the EMB benchmark. Our evaluation results indicate that MISH performs comparably to MOSA, a state-of-the-art technique for generating test cases for microservices, and outperforms MOSA in certain scenarios. MISH achieves this performance while utilizing only a single objective to determine fitness for test cases, whereas MOSA utilizes multiple objectives. MISH provides promising results for the use of model inference as a search heuristic in test case generation. Furthermore, our results suggest that MISH is quicker at discovering unseen targets at the start of the search compared to MOSA.  

In future work, we plan to investigate the effect of utilizing MISH as part of a many-objective search algorithm (e.g., combining MOSA and MISH). We speculate that by incorporating MISH within a many-objective search algorithm, we can aid the many-objective search algorithm in discovering unseen targets quicker at the start of the search. Moreover, we can utilize different information gathered by the many-objective search algorithm to help MISH with its search when its performance reaches a plateau.

In its current implementation, MISH communicates with the model inference framework via file-based interactions, which involve frequent disk I/O operations. Over many generations of test cases, this file-based communication negatively impacts performance. To address this limitation, we plan to replace the file-based communication with a specialized API. This new communication mechanism would not only improve efficiency but also enhance the robustness of MISH by mitigating potential communication failures.
\balance
\bibliographystyle{plain}
\bibliography{references}
\end{document}